\documentclass[aps,prx,twocolumn,showpacs,superscriptaddress]{revtex4-2}

\usepackage{amsmath}
\usepackage[dvipdfmx]{graphicx}
\usepackage{siunitx}
\usepackage[version=4]{mhchem}
\usepackage{color}

\usepackage{url}
\usepackage[colorlinks=true, allcolors=blue]{hyperref}
\usepackage{hyperref}

\begin{document}

\title{Optimal elemental configuration search in crystal using quantum approximate optimization algorithm}

\author{Kazuhide Ichikawa}
\email{ichikawa.kazuhide@jp.panasonic.com}
\affiliation{Technology Sector, Panasonic Holdings Corporation, 1006 Kadoma, Kadoma City, Osaka 571-8508, Japan}

\author{Genta Hayashi}
\email{hayashi.genta@jp.panasonic.com}
\affiliation{Technology Sector, Panasonic Holdings Corporation, 1006 Kadoma, Kadoma City, Osaka 571-8508, Japan}

\author{Satoru Ohuchi}
\email{ohuchi.s@jp.panasonic.com}
\affiliation{Technology Sector, Panasonic Holdings Corporation, 1006 Kadoma, Kadoma City, Osaka 571-8508, Japan}

\author{Tomoyasu Yokoyama}
\email{yokoyama.tomoyasu@jp.panasonic.com}
\affiliation{Technology Sector, Panasonic Holdings Corporation, 1006 Kadoma, Kadoma City, Osaka 571-8508, Japan}

\author{Ken N. Okada}
\email{okada.ken.qiqb@osaka-u.ac.jp}
\affiliation{Center for Quantum Information and Quantum Biology, Osaka University, 1-2 Machikaneyama, Toyonaka, Osaka 560-0043, Japan}

\author{Keisuke Fujii}
\email{fujii@qc.ee.es.osaka-u.ac.jp}
\affiliation{Center for Quantum Information and Quantum Biology, Osaka University, 1-2 Machikaneyama, Toyonaka, Osaka 560-0043, Japan}
\affiliation{School of Engineering Science, Osaka University, 1-3 Machikaneyama, Toyonaka, Osaka 560-8531, Japan}
\affiliation{RIKEN Center for Quantum Computing (RQC), 2-1 Hirosawa, Wako, Saitama 351-0198, Japan}

\date{\today}

\begin{abstract}
Optimal elemental configuration search in crystal is a crucial task to discovering industrially important materials such as lithium-ion battery cathodes.
In this paper we present application of quantum approximate optimization algorithm, the representative near-term quantum algorithm for combinatorial optimization, to finding the most stable elemental configuration in a crystal, using Au-Cu alloys as an example.
After expressing the energy of the crystal in the form of the Ising model through the cluster expansion method combined with first-principles calculations, we numerically perform QAOA with three types of parameter optimization. 
As a result, we have demonstrated that the optimal solution can be sampled with a high probability for crystals containing up to 32 atoms.
Our results could pave the way for optimal elemental configuration search in a crystal using a near-term quantum computer.

\end{abstract}


\maketitle

\section{Introduction}

Development of new materials increasingly relies on computational methods to predict and understand their physical properties \cite{louie2021discovering, bartel2022review,marzari2021electronic}.
First-principles calculations are essential for this task, yet they require detailed knowledge of the atomic structure of the target material, which is often incomplete, particularly in multinary crystal systems, where multiple types of atoms can occupy various lattice sites. For example, in lithium-ion battery cathodes \cite{booth2021perspectives} like \( \ce{LiMO2} \) (M = Ni, Co, Mn), while the overall layered structure is known, the exact positions of the transition metals within the layers remain uncertain. 
Because the configuration of these metals directly influences battery capacity and potential, accurately determining their site positions is vital for optimizing material performance. Such methodologies are also increasingly needed for other functional materials in environmental and energy applications, including batteries, capacitors, and catalysts \cite{ma2021high,sturman2022high,schweidler2024high,reddy2013metal,nzereogu2022anode,xu2021garnet,aravindan2018building,sun2021high}.

Identifying stable crystal configurations involves finding out the lowest energy arrangement of atoms, a process we refer to as the elemental configuration optimization problem.
This problem is challenging due to the high computational cost of first-principles calculations and the exponentially large number of possible atomic configurations.
Traditionally, techniques like the ``cluster expansion method" \cite{SANCHEZ1984334, de1994solid, ceder2000first} to reduce computational cost of crystal energy and simulated annealing to solve the combinatorial optimization problem \cite{ozolicnvs1998cu, seko2006prediction,chang2019clease,yang2020arrangement} have been employed to address these issues, enabling more efficient exploration of stable crystal structures.
In recent years, there has been renewed interest in the fact that the crystal energy function derived from the cluster expansion method naturally takes the form of an Ising-type function, where elemental arrangements are represented by discrete variables.
Along with the advancement of Ising machines \cite{mohseni2022ising}, which are dedicated devices designed to efficiently solve the Ising model, research—including that by some of the authors—has begun to explore the optimization of the cluster expansion method using Ising machines \cite{choubisa2023accelerated,mendoza2023optimization,ichikawa2024accelerating}. Additionally, there are studies utilizing Ising machines for crystal structure optimization with energy functions derived from methods other than the cluster expansion method \cite{binninger2024optimization,couzinie2024annealing,gusev2023optimality,couzinie2024machine}.

Recently, the development of intermediate-scale quantum computers, known as noisy intermediate-scale quantum (NISQ) devices \cite{preskill2018quantum}, 
has been remarkable, and there have been intensive studies on their applications for combinatorial optimization algorithms \cite{preskill2018quantum}.
One of the most promising approaches is a hybrid quantum-classical algorithm called quantum approximate optimization algorithm (QAOA) \cite{farhi2014quantum},
where a parameterized quantum circuit is employed to search a solution via a variational approach \cite{cerezo2021variational}.
Contrast to the Ising machines using simulated annealing or quantum annealing machines, 
QAOA can fully leverage the programmability of a universal quantum computer. 
Specifically, instead of introducing soft constraints via penalty terms, 
QAOA can search a solution whithin a subspace satisfying the constraints~\cite{PhysRevA.101.012320,PhysRevResearch.5.023071}.
It can also handle higher-order Ising models (higher-order unconstrained binary optimization: HUBO)~\cite{pelofske2024short}.
Unfortunately, so far, QAOA has been mainly studied on generic combinatorial optimization problems including the MaxCut problem and the low autocorrelation binary sequence (LABS) problem \cite{harrigan2021quantum,shaydulin2024evidence,montanez2024towards}. 
On the other hand, while there are a few works on application of QAOA to molecular design \cite{gao2023quantum,gao2023combined}, its application in the field of materials science remains largely unexplored.

\begin{figure*}[ht]
\centering
\includegraphics[width=0.95\textwidth]{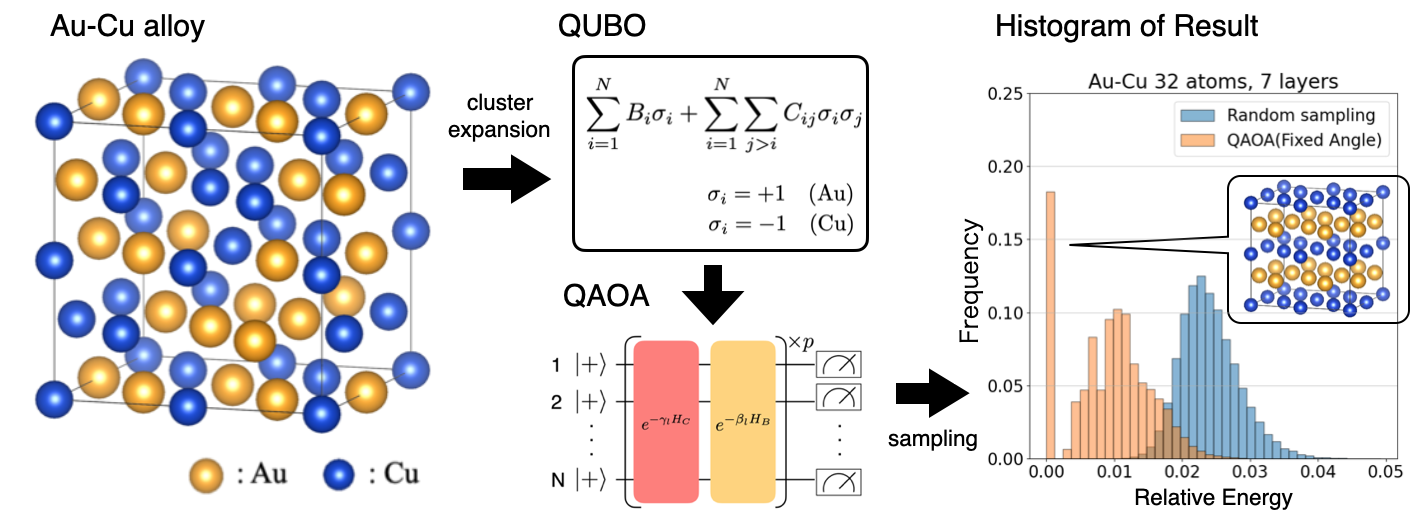}
\caption{Schematic diagram presenting overview of our QAOA-based elemental configuration optimization for the Au-Cu alloy.
First, the energy of the Au-Cu crystal structure is expressed in a QUBO form using the cluster expansion method. 
The QUBO variables correspond to the atomic sites in the crystal, where we set 
\(\sigma_{i} = +1\) if site \(i\) is occupied by Au and \(\sigma_{i} = -1\) if it is occupied by Cu.
Next, this QUBO energy function is optimized by the QAOA algorithm.
VESTA \cite{momma2011vesta} was used for the visualization of the crystal structures in this figure.}
\label{fig:g_abst}
\end{figure*}

In this study, we explore application of QAOA to the optimal elemental configuration search in a crystal.
Taking up Au-Cu alloys as an example, we solve the Ising model obtained from the cluster expansion method combined with first-principle calculations using QAOA with full parameter optimization and fixed-angle approaches (a schematic diagram of our research is shown in Fig.~\ref{fig:g_abst}).
Here, the “fixed angle approach” refers to a method of presetting QAOA parameters in a way that does not depend on individual problem instances \cite{brandao2018fixed}, which provides a significant advantage when running QAOA on real quantum hardware.
Using Qulacs, a high-speed quantum computer simulator \cite{suzuki2021qulacs},
we conducted simulations with up to 32 qubits and demonstrated that QAOA can find the optimal configuration with high probability.
We also examined how performance varies under different scenarios including using pre-optimized fixed angles and transferring parameters optimized with smaller-scale problems.
We find out that the bit string corresponding to the optimal configuration can be sampled with a high probability even when using fixed-angle parameters.
This is an important observation because when considering implementation on real NISQ devices it is necessary to use parameters determined in advance because full parameter optimization is difficult for a large system because of high sampling cost and hardware error.
Our findings are expected to be the first step forward towards optimizing elemental configurations in a large crystalline system.

The rest of the paper is organized as follows. In Section~\ref{sec:methods}, we briefly explain the cluster expansion method and QAOA, and describe how the energy function can be obtained in a format suitable for optimization by QAOA. In Section~\ref{sec:results}, we present the calculation results using the Cu-Au alloy as an example and demonstrate that the minimum energy configuration is successfully obtained for crystal systems of up to 32 atoms using QAOA implemented on a simulator. Finally, we provide a summary and discuss future prospects in Section~\ref{sec:conclusion}.

\section{Methods}  \label{sec:methods}

\subsection{Cluster Expansion Method}  \label{sec:CE}

In this subsection, we briefly outline the cluster expansion method as used in computational materials science. We focus specifically on how the cluster expansion method represents the energy of crystal structures in a format suitable for optimization by QAOA. For a more detailed explanation, readers are encouraged to consult references \cite{SANCHEZ1984334, de1994solid, ceder2000first, wu2016cluster, barroso2022cluster} or explore available program packages that implement the cluster expansion method \cite{van2009multicomponent, seko2009cluster, troppenz2017predicting, chang2019clease, aangqvist2019icet, puchala2023casm}. 

The cluster expansion method leverages fixed positions of potential atomic sites, allowing the energy to be expressed in a way that facilitates fast calculation. 
Based on the fixed lattice structure of the crystal, it determines the possible atomic sites. Instead of representing \(N\) atomic positions with \(N\) real-valued vectors, it uses an \(N\)-dimensional configuration vector \( \vec{\sigma} \) of discrete variables corresponding to atomic species to describe the energy \( E(\vec{\sigma}) \).
The energy is expanded in terms of basis functions \( \varphi_\alpha(\vec{\sigma}) \) as:
\begin{eqnarray}
E(\vec{\sigma}) = \sum_\alpha V_\alpha \varphi_\alpha(\vec{\sigma}),   \label{eq:E_CE}
\end{eqnarray}
where \( V_\alpha \) are the expansion coefficients fitted to energies obtained from first-principles calculations. These coefficients, learned from small crystal structures, can be applied to larger systems, making the method highly transferable.

For simplicity, this paper considers a binary crystal structure consisting of \( N \) atomic sites, approximated up to the second term in the expansion:
\begin{eqnarray}
E(\vec{\sigma}) = A + \sum_{i=1}^N B_i \sigma_i + \sum_{i=1}^N \sum_{j>i} C_{ij} \sigma_i \sigma_j, \label{eq:ce_2nd}
\end{eqnarray}
where \( \sigma_i = \pm 1 \), yielding an energy function akin to the Ising model. This form is suitable for optimization using QAOA. 
In the following sections, we apply this approach to the Au-Cu alloy to demonstrate its effectiveness.

Here, we summarize the computational methods and conditions employed in this study to obtain coefficients in Eq.~\eqref{eq:ce_2nd}, which are consistent with those used in our previous work \cite{ichikawa2024accelerating}. However, they are reiterated here for the sake of self-containedness and convenience. The cluster expansion calculations to determine \( V_\alpha \) in Eq.~\eqref{eq:E_CE} were carried out using the Alloy Theoretic Automated Toolkit (ATAT) \cite{van2009multicomponent}, while the first-principles DFT calculations were performed using the Vienna Ab Initio Simulation Package (VASP) \cite{kresse1993ab,kresse1994ab,kresse1996efficiency,kresse1999ultrasoft}. For the VASP calculations, we employed the Projector Augmented Wave method with the Perdew-Burke-Ernzerhof functional \cite{perdew1996generalized}. The energy cutoff was set at 355 eV, and the \( k \)-point mesh was generated with a 0.02\,\AA\(^{-1}\) interval, including the \( \Gamma \) point. The lattice constants and atomic positions were relaxed, with a convergence criterion of 0.02\,eV\,{\AA}\(^{-1}\) for the forces. The base cell for the cluster expansion method was a face-centered cubic (fcc) lattice with a side length of 3.8\,{\AA}, containing four atoms. 
The cluster dimension was restricted to the second order, with a size constraint of 7.6\,{\AA} or smaller. Under these conditions, the largest training dataset for the cluster expansion method corresponded to a supercell containing 14 atoms.
In Sec.~\ref{sec:results}, we refer to the crystal systems with numbers of atoms 12, 16, 24, and 32, which correspond to supercells of sizes $1\times 1\times 3$,  $1\times 2\times 2$,  $1\times 2\times 3$, and $2\times 2\times 2$, respectively.

\subsection{Quantum Approximation Optimization Algorithm}   \label{sec:QAOA}

In this subsection, we briefly explain QAOA to set up our notation. For a more detailed explanation, we refer to recent review papers \cite{mohseni2022ising, blekos2024review}.

QAOA is a hybrid quantum-classical algorithm for solving combinatorial optimization problems with the cost function expressed in the form of the Ising model.
First of all, our cost Hamiltonian is in the following form of the Ising model with the coupling constants $h_i$ and $J_{ij}$:
\begin{eqnarray}
H_C = \sum_{i=1}^N  h_i Z_i + \sum_{i=1}^N \sum_{j>i} J_{ij} Z_i Z_j,  \label{eq:Hcost}
\end{eqnarray}
where $N$ is the number of qubits, $Z_i$ is the Pauli-$Z$ operator acting on the $i$-th qubit and the constant term is omitted because it does not affect the QAOA formulation.
Next, we define the mixer Hamiltonian
\begin{eqnarray}
H_B = \sum_{i=1}^N X_i,
\end{eqnarray}
where $X_i$ is the Pauli-$X$ operator acting on the qubit $i$.
Then, defining the following operator
\begin{eqnarray}
U(H, \alpha) = e^{-i \alpha H}, 
\end{eqnarray}
the quantum circuit of QAOA can be written as follows:
\begin{eqnarray}
U_{\rm QAOA}(\vec{\gamma}, \vec{\beta}) =  \prod_{l=1}^{p} U(H_B, \beta_l)  U(H_C, \gamma_l), 
\end{eqnarray}
which is parametrized by $2p$ rotation angles, $\vec{\gamma} = (\gamma_1, \gamma_2, ..., \gamma_p)$ and $\vec{\beta} = (\beta_1, \beta_2, ..., \beta_p)$.
Finally, the QAOA ansatz state $| \vec{\gamma}, \vec{\beta} \rangle$ is prepared as
\begin{eqnarray}
| \vec{\gamma}, \vec{\beta} \rangle = U_{\rm QAOA}(\vec{\gamma}, \vec{\beta})  | + \rangle^{\otimes N},  \label{eq:stateQAOA}
\end{eqnarray}
where $ | + \rangle^{\otimes N}$ is the eigenstate of the mixer Hamiltonian $H_B$ with the largest eigenvalue (the lowest energy state of the Hamiltonian $-H_B$).

In QAOA, we expect that the optimal solution for the classical cost function
$C(\vec{\sigma} ) = \sum_{i=1}^N  h_i \sigma_i + \sum_{i=1}^N \sum_{j>i} J_{ij} \sigma_i \sigma_j$, 
where \( \sigma_i = \pm 1 \), or nearly optimal solution to it, can be heuristically obtained by sampling the ansatz state given in Eq.~\eqref{eq:stateQAOA} with high probability,
provided that the rotation angles $(\vec{\gamma}, \vec{\beta})$ are chosen appropriately. 
In the original QAOA paper \cite{farhi2014quantum}, it is proposed that $(\vec{\gamma}, \vec{\beta})$
 be chosen to optimize the following QAOA cost function: 
\begin{eqnarray}
F_p(\vec{\gamma}, \vec{\beta}) = \langle \vec{\gamma}, \vec{\beta} | H_C | \vec{\gamma}, \vec{\beta} \rangle,  \label{eq:costQAOA}
\end{eqnarray}
which is the expectation value of the cost Hamiltonian $H_C$ (Eq.~\eqref{eq:Hcost}) with respect to the ansatz state (Eq.~\eqref{eq:stateQAOA}).
This optimization process involves calculating the expectation value represented by Eq.~\eqref{eq:costQAOA} on a quantum computer, 
while updating the $2p$ hyperparameters $(\vec{\gamma}, \vec{\beta})$ is performed on a classical computer, resulting in a hybrid approach.
Such optimization process becomes increasingly difficult as the number of variables in the cost function increases, making it more likely to get trapped in local minima. This has been considered a drawback of QAOA.
One idea to address this issue is the fixed-angle QAOA \cite{brandao2018fixed,galda2021transferability}, where parameters determined from a smaller qubit problem are directly applied to a larger problem  \cite{wurtz2021fixed}. 
Additionally, it is commonly tried recently to reduce the number of parameters by restricting $(\vec{\gamma}, \vec{\beta})$ to depend linearly on 
the depth of the layers 
(resulting in only four parameters: the slopes and intercepts for $\vec{\gamma}$ and $\vec{\beta}$) \cite{sakai2024linearly,montanez2024towards}. 
There could also be methods that combine these approaches for parameter determination. 
In this paper, we will test and compare several of these methods.

The expectation value computation of Eq.~\eqref{eq:costQAOA} and sampling of the QAOA ansatz state Eq.~\eqref{eq:stateQAOA} are performed in this paper by the quantum circuit simulator Qulacs \cite{suzuki2021qulacs}, that is, by an ideal simulator without any noise source.

\begin{figure*}[ht]
\centering
\includegraphics[width=0.95\textwidth]{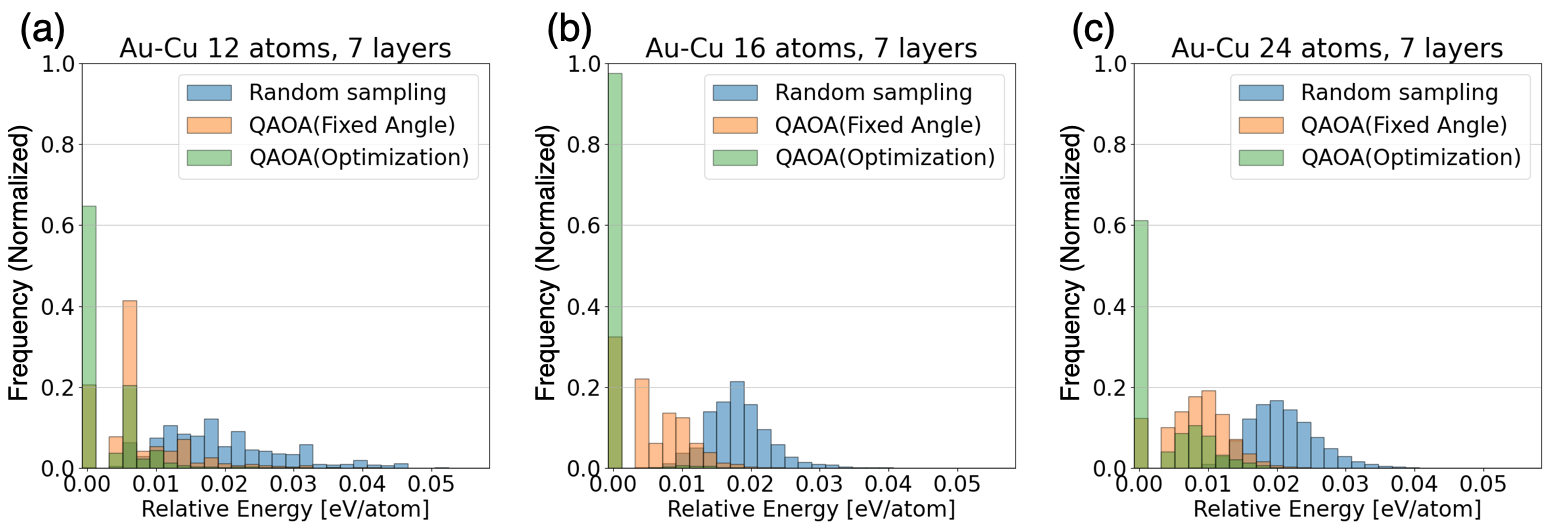}
\caption{Frequency distribution of relative energies of the Au-Cu alloy sampled using three different methods—random sampling, QAOA with fixed angles, and QAOA with optimization. QAOA employs 7 layers, and the Au-Cu systems are presented for (a) 12 atoms, (b) 16 atoms, and (c) 24 atoms.}
\label{fig:hist_compare}
\end{figure*}

\section{Numerical Results}   \label{sec:results}

In this study, the Au-Cu binary alloy was selected as a representative binary crystal structure. This material system is the same as that used in the authors' previous work \cite{ichikawa2024accelerating}, and the cluster expansion calculations were performed using the same methods as described in Section~\ref{sec:CE} to derive the coefficients in Eq.~\eqref{eq:ce_2nd}. 

To optimize the cluster expansion energy function using QAOA, we set the coefficients of the Ising model to be \( J_{ij}=C_{ij} \) and \( h_i=B_i \) from the comparison of Eqs.~\eqref{eq:ce_2nd} and \eqref{eq:Hcost}.
Note that both \( C_{ij} \) and \( B_i \) are real numbers and are not restricted to integers. The number of qubits used in QAOA corresponds to the number of atoms in the targeted crystal system.

In the following sections, 
we take three different approaches to set the parameters \( F_p(\vec{\gamma}, \vec{\beta}) \) for QAOA:
full parameter optimization (Sec.~\ref{sec:res_QAOA}) taken in the original method by Farhi et al.\cite{farhi2014quantum}, 
parameter transfer (Sec.~\ref{sec:res_transfer}) as a means to bypass the optimization process in the original QAOA method,
and optimization of parameters under linear constraint (Sec.~\ref{sec:res_linear}), where the parameters are made to depend linearly on the depth of the QAOA layers, an approach referred to as the ``linear ramp QAOA."

\subsection{QAOA}    \label{sec:res_QAOA}
Firstly, we follow the original method proposed by Farhi et al.~\cite{farhi2014quantum} to optimize the QAOA cost function \( F_p(\vec{\gamma}, \vec{\beta}) \) (Eq.~\eqref{eq:costQAOA}) by searching for suitable rotation angle parameters \((\vec{\gamma}, \vec{\beta})\). These optimized parameters are then used to perform QAOA sampling. In this study, the number of sampling iterations for all cases is set to 100,000. The optimization of the rotation angles was conducted using the Broyden-Fletcher-Goldfarb-Shanno (BFGS) algorithm.

As an example of the computational results, Fig.~\ref{fig:hist_compare} shows the results of solving the elemental configuration optimization problem for Au-Cu alloys with 12, 16, and 24 atoms using a 7-layer QAOA. In Fig.~\ref{fig:hist_compare}, the horizontal axis represents relative energy values (compared to the minimum energy obtained through exhaustive search), and the vertical axis indicates the frequency (normalized by the number of samples), depicted as a histogram. The data labeled as ``QAOA (Optimization)" represent sampling results after optimizing the rotation angles, whereas those labeled as ``QAOA (Fixed-Angle)" come from sampling with pre-fixed rotation angles, and those labeled as ``Random sampling” are generated by purely random sampling. 

The ``QAOA (Optimization)" data show a very high proportion of samples at the minimum energy, demonstrating the success of QAOA in identifying the minimum-energy configuration. Compared to ``QAOA (Fixed-Angle),” the frequency of sampling the minimum energy is higher for ``QAOA (Optimization),” confirming the effectiveness of the QAOA method proposed in Ref.~\cite{farhi2014quantum}. We note in passing that, in these histograms, the leftmost bin contains only the minimum energy, and there is a gap before the next bin, reflecting the relatively large energy gap between the configuration yielding the minimum energy and the next-lowest-energy configuration for these crystal systems.

\begin{figure}[ht]
\centering
\includegraphics[width=0.95\columnwidth]{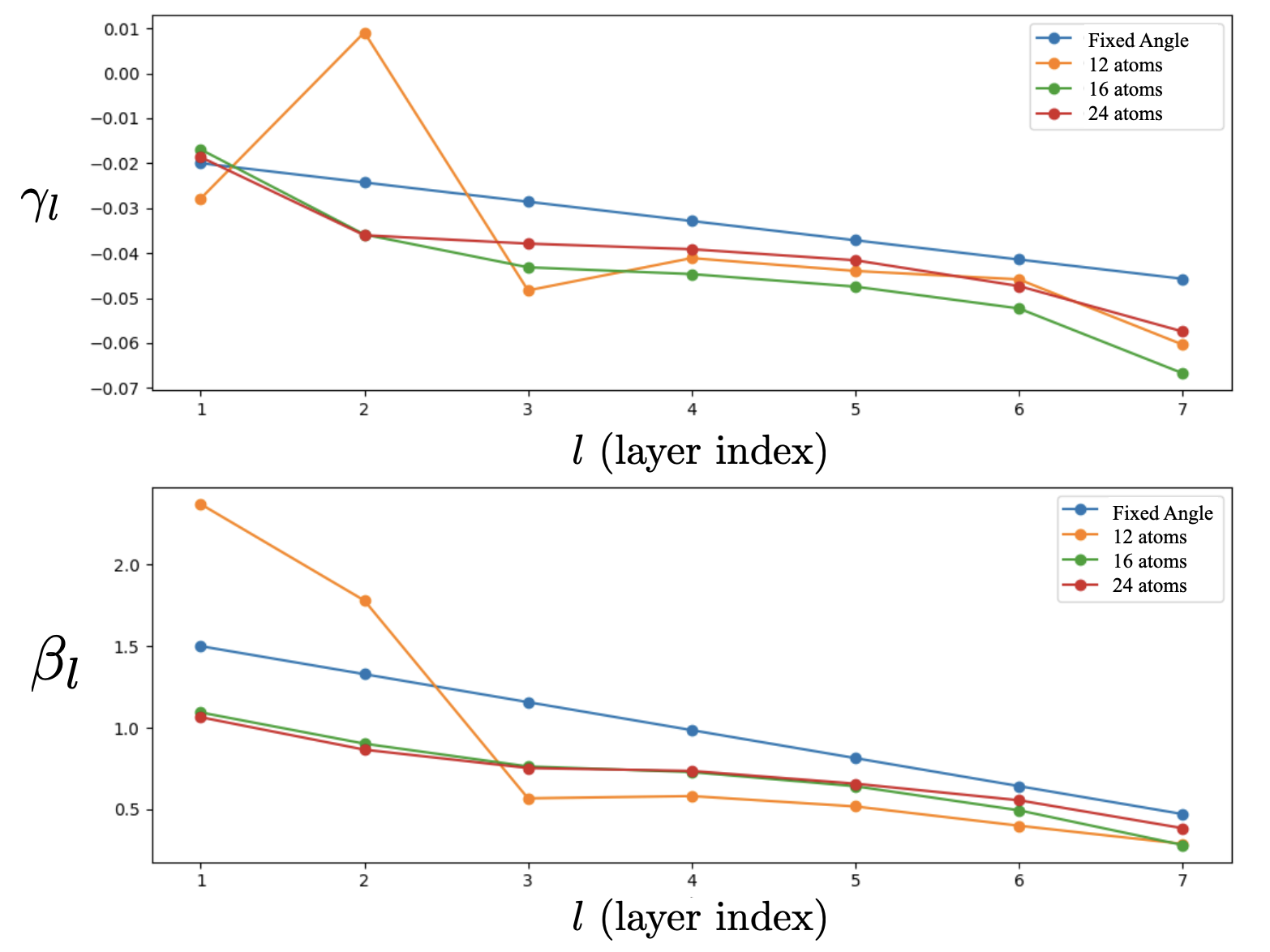}
\caption{Comparison of rotation angle parameters before optimization (Fixed Angle) and after optimization for systems with 12, 16, and 24 atoms.}
\label{fig:opt_angles}
\end{figure}

Here, the term ``Fixed Angle" refers to the initial values of the rotation angles used in the optimization process. These values are explicitly defined and are designed to depend linearly on the depth of the layers as follows:
\begin{eqnarray}
 \gamma_l &=&  -b_I - \frac{\alpha_I}{p} (l - 1) \label{eq:gamma_linear} \\
 \beta_l &=&  b_z - \frac{\alpha_z}{p} (l - 1)  \label{eq:beta_linear}
\end{eqnarray}
where the parameters are set to \(\alpha_I = 0.03\), \(b_I = 0.20\), \(\alpha_z = 1.2\), and \(b_z = 1.5\). These parameter values have been shown to yield effective solutions in fully connected Ising models with random coefficients involving approximately up to 10 qubits \cite{sakai2024linearly}. For this study, the values of \(\gamma\) were scaled to fit the range of the cost function used in this problem.

Figure~\ref{fig:opt_angles} illustrates the Fixed Angle parameters alongside the optimized rotation angle parameters. 
As seen from the figure, for the case of 12 atoms, the optimized parameters deviate significantly from the initial linear values, 
whereas for the cases of 16 and 24 atoms, the (sub)optimal values of the parameters are obtained close to the initial values, given 
as linear in the depth of the QAOA layers.

\begin{table}[htp]
\caption{Success rates (\%) of QAOA with optimized rotation angles. Results are shown for QAOA ansatz with layer numbers \(p = 5 \sim 9\).}
\begin{center}
\begin{tabular}{|c|c|c|c|}
\hline
Sampling  & 12 atoms & 16 atoms & 24 atoms \\
 \hline
 \hline
 $p=5$   & 47.4  & 82.1  & 34.4  \\
 $p=6$   & 53.6  & 95.2  & 37.6  \\
 $p=7$   & 64.8  & 97.4  & 61.2  \\
 $p=8$   & 71.4  & 96.1  & 69.5  \\
 $p=9$   & 85.2  & 97.9  & 71.2  \\
\hline
\end{tabular}
\end{center}
\label{tab:resQAOA_opt}
\end{table}%

\begin{table}[htp]
\caption{Success rates (\%) of QAOA with fixed angles (initial values for optimization). Results are shown for QAOA ansatz with layer numbers \(p = 5 \sim 9\).}
\begin{center}
\begin{tabular}{|c|c|c|c|c|}
\hline
Sampling  & 12 atoms & 16 atoms & 24 atoms & 32 atoms \\
 \hline
 \hline
 $p=5$  & 11.7  & 13.7  & 2.1  & 2.1 \\
 $p=6$  & 15.1  & 28.6  & 7.8 & 9.8  \\
 $p=7$  & 20.5  & 32.4  & 12.2 & 18.6  \\
 $p=8$  & 14.2  & 33.1  & 18.4 & 29.0  \\
 $p=9$  & 29.5  & 37.7  & 21.5 & 27.6  \\
\hline
\end{tabular}
\end{center}
\label{tab:resQAOA_fixed_angle}
\end{table}%

\begin{table*}[htp]
\caption{Success rates (\%) achieved through parameter transfer. The first row indicates the number of atoms used to optimize the rotation angle parameters in the cost function, while the second row represents the number of atoms in the systems where QAOA sampling was performed using these optimized parameters. Improvements over the fixed-angle results in Table~\ref{tab:resQAOA_fixed_angle} are highlighted in bold.}
\centering
\begin{tabular}{|c|c|c|c|c|c|c|}
\hline
Optimization & \multicolumn{3}{|c|}{12 atoms } & \multicolumn{2}{|c|}{16 atoms } & 24 atoms \\
\hline
Sampling  & 16 atoms & 24 atoms & 32 atoms & 24 atoms & 32 atoms & 32 atoms \\
\hline
\hline
 $p=5$  & {\bf 62.2} & {\bf 15.4} & 0.3  & {\bf 26.6} & 0.67  & {\bf 6.5}  \\
 $p=6$  & {\bf 57.7} & {\bf 13.1} & 0.29 & {\bf 39.0} & {\bf 11.5}  & 6.6  \\
 $p=7$  & {\bf 50.1} & 8.1  & 1.3  & {\bf 43.2} & 14.1  & 13.6 \\
 $p=8$  & {\bf 42.2} & 6.2  & 1.0  & {\bf 44.1} & 14.2  & 18.1 \\
 $p=9$ & 25.6 & 1.6  & 4.9  & {\bf 45.7} & 15.5  & 17.5 \\
\hline
\end{tabular}
\label{tab:resQAOA_transfer}
\end{table*}

\begin{table*}[htbp]
\caption{Success rates (\%) achieved through parameter transfer in Linear Ramp QAOA. The first row indicates the number of atoms used for cost function optimization of the rotation angle parameters, and the second row represents the number of atoms in the systems where QAOA sampling was performed using the optimized parameters. Improvements over the fixed-angle results in Table~\ref{tab:resQAOA_fixed_angle} are highlighted in bold.}
\centering
\begin{tabular}{|c|c|c|c|c|c|c|c|c|c|}
\hline
Optimization & \multicolumn{4}{|c|}{12 atoms } & \multicolumn{3}{|c|}{16 atoms } &  \multicolumn{2}{|c|}{24 atoms } \\
\hline
Sampling  & 12 atoms & 16 atoms & 24 atoms & 32 atoms & 16 atoms & 24 atoms & 32 atoms & 24 atoms & 32 atoms \\
\hline
\hline
 $p=5$  & {\bf 35.9} & {\bf 46.4} & {\bf 9.1}  & 0.2 & {\bf 80.0}  & {\bf 26.7}  & 0.6   & {\bf 26.8}  & {\bf 5.0} \\
 $p=6$  & {\bf 39.4} & {\bf 50.4} & 7.7  & 1.4 & {\bf 81.7}  & {\bf 31.2}  & 0.7  & {\bf 30.8}  & 2.2\\
 $p=7$  & {\bf 64.7} & {\bf 54.0}  & 5.7 & 17.1  & {\bf 84.3} & {\bf 29.2} & 0.2 & {\bf 30.5} & 1.0  \\
 $p=8$  & {\bf 67.7} & {\bf 54.1} & 6.2  & 18.1  & {\bf 84.9}  & {\bf 30.3}  & 0.7  & {\bf 27.9}  & 5.3\\
 $p=9$  &  {\bf 72.6} & 25.2 & 3.2  & 5.1 & {\bf 82.8}  & {\bf 32.6}  & 3.5  & {\bf 39.5}  & 4.2 \\
\hline
\end{tabular}
\label{tab:resQAOA_linear}
\end{table*}

\begin{table*}[htbp]
\caption{Success rates (\%) achieved through parameter transfer in Linear Ramp QAOA with $\beta$ fixed and only $\gamma$ optimized. The first row indicates the number of atoms used for cost function optimization of the rotation angle parameters, and the second row represents the number of atoms in the systems where QAOA sampling was performed using the optimized parameters. Improvements over the fixed-angle results in Table~\ref{tab:resQAOA_fixed_angle} are highlighted in bold.}
\centering
\begin{tabular}{|c|c|c|c|c|c|c|c|c|c|}
\hline
Optimization & \multicolumn{4}{|c|}{12 atoms } & \multicolumn{3}{|c|}{16 atoms } &  \multicolumn{2}{|c|}{24 atoms }   \\
\hline
Sampling  & 12 atoms & 16 atoms & 24 atoms & 32 atoms & 16 atoms & 24 atoms & 32 atoms & 24 atoms & 32 atoms \\
\hline
\hline
 $p=5$  & {\bf 14.2}  & {\bf 14.6}  & {\bf 3.3}   & {\bf 4.4}   & {\bf 55.2} & {\bf 12.9}  & {\bf 2.5} &  {\bf 12.5} & {\bf 8.3}  \\
 $p=6$  & {\bf 19.3}  & 21.9  & 5.8   & 8.2  &  {\bf 67.6}  & {\bf 17.9}  & 1.3   & {\bf 19.4}   & 8.1  \\
 $p=7$  & {\bf 26.1} & {\bf 44.8}  & {\bf 15.2} & 14.4 & {\bf 78.0}  & {\bf 22.7} & 0.01 & {\bf 25.6} & 4.3 \\
 $p=8$  & {\bf 34.1} & {\bf 61.8} & {\bf 24.0}  & 12.0  & {\bf 79.4}   & {\bf 25.5}   & 3.6   & {\bf 31.8}   & 1.8  \\
 $p=9$  & {\bf 38.6}  & {\bf 71.8}  & {\bf 30.1}  & 4.7  & {\bf 78.0}  & {\bf 25.2}   & 0.9   & {\bf 30.2}   &  1.6 \\
\hline
\end{tabular}
\label{tab:resQAOA_linear2}
\end{table*}

\begin{table}[htp]
\caption{Success rates (\%) achieved through parameter transfer in Linear Ramp QAOA with $\beta$ fixed and only $\gamma$ optimized. The $\gamma$ optimization was performed using the cost function for the 32-atom system, and the optimized values were applied to QAOA sampling. The first row indicates the number of atoms in the systems where QAOA sampling was performed. Improvements over the fixed-angle results in Table~\ref{tab:resQAOA_fixed_angle} are highlighted in bold.}
\begin{center}
\begin{tabular}{|c|c|c|c|c|}
\hline
Sampling  & 12 atoms & 16 atoms & 24 atoms & 32 atoms \\
 \hline
 \hline
 $p=5$  & {\bf 19.8}  & {\bf 31.3}  & {\bf 9.2}  & {\bf 10.1} \\
 $p=6$  & {\bf 20.7}  & {\bf 36.5}  & {\bf 12.5} & {\bf 16.7}  \\
 $p=7$  & {\bf 25.1}  & {\bf 38.2}  & {\bf 15.7} & {\bf 24.1}  \\
 $p=8$  & {\bf 17.0}  & 31.2  & 14.8 & {\bf 35.7}  \\
 $p=9$  & 29.3  & {\bf 41.9}  & 18.4 & {\bf 34.1}  \\
\hline
\end{tabular}
\end{center}
\label{tab:resQAOA_linear3}
\end{table}%

Similar QAOA calculations were conducted for layer numbers \(p = 5, 6, 8, 9\). As in the case of \(p = 7\), it was observed that the proportion of samples achieving the minimum energy (hereafter referred to as the ``success rate") was enhanced. The results of these calculations are summarized in Table~\ref{tab:resQAOA_opt}. In each case, the success rate was significantly higher than that obtained with ``Fixed Angle" parameters (shown in Table~\ref{tab:resQAOA_fixed_angle}). 

A general trend observed in Table~\ref{tab:resQAOA_opt} is that, for the same number of atoms, the success rate increases with larger \(p\). This is a reasonable outcome, as the representational capacity of the QAOA ansatz improves with the number of layers \(p\). However, a closer look reveals that, for 16 atoms, the success rate for \(p = 8\) is slightly lower than that for \(p = 7\). 
This can be attributed to the fact that, for 16 atoms, the success rate already approaches 100\% at \(p = 6\), making it exceedingly challenging to further increase the success rate through parameter optimization as \(p\) grows larger. For \(p = 8\) in the case of 16 atoms, the optimization did not perform well, failing to fully leverage the increased representational capacity.

Additionally, similar calculations were attempted for 32 atoms. However, due to the high computational cost of expectation value calculations in the state vector simulation used in this study, the BFGS optimization did not converge even for \(p = 5\).

\subsection{Parameter Transfer}   \label{sec:res_transfer}

As mentioned at the end of the previous subsection, the optimization of rotation angle parameters becomes increasingly challenging as the number of atoms (and therefore qubits) increases, as well as with a larger number of layers. To efficiently solve larger systems, it is necessary to avoid this parameter optimization process wherever possible. In this subsection, we explored the concept of fixed-angle QAOA, as described in Section~\ref{sec:QAOA}. Specifically, we attempted to transfer the optimized rotation angle parameters obtained from the cost function of smaller systems to the QAOA sampling for larger systems. 
An important advantage of parameter transfer is that, by circumventing the parameter optimization step using quantum computers, 
one can reduce quantum computational cost significantly, which is crucial for a feasible implementation of large-scale problems.

The results of this approach are summarized in Table~\ref{tab:resQAOA_transfer}. When the optimized rotation angles from the cost function of a 12-atom system were applied to the sampling of a 16-atom system, the success rate, while naturally lower than the success rate achieved by directly optimizing the cost function for 16 atoms (as shown in the previous subsection), remained relatively high, indicating that parameter transfer was successful. Similarly, when the parameters optimized for 16 atoms were transferred to the sampling of 24 atoms, the results were also favorable. 

On the other hand, transferring the parameters optimized for 12 atoms to the sampling of 24 atoms resulted in lower success rates, demonstrating limited effectiveness. This is likely due to overfitting during the optimization of the cost function for 12 atoms, leading to lower generalization performance.

For the 32-atom system, the success rate was low but still finite, a result unattainable through random sampling. However, when comparing the ``32 atoms" column in Table~\ref{tab:resQAOA_fixed_angle} with that in Table~\ref{tab:resQAOA_transfer}, the success rates were generally higher in the former (highlighted in bold in Table~\ref{tab:resQAOA_fixed_angle} for cases where improvement was achieved). This suggests that parameter transfer in these cases led to a reduction in success rates.

It is important to note that the number of atoms considered in this study is limited by the computational constraints of using a simulator. Therefore, it is premature to conclude that the method of transferring optimal parameters from smaller systems to larger systems is ineffective. For larger-scale computations, particularly when utilizing real quantum hardware, the results could potentially differ.

\subsection{Linear Ramp QAOA}   \label{sec:res_linear}

In the previous subsection, we pointed out the possibility of overfitting and attempted to address this issue by constraining the functional form of the rotation angle parameters. The functional form used here is the same as the one employed for specifying the initial values of optimization, given by Eqs.~\eqref{eq:gamma_linear} and \eqref{eq:beta_linear}, which are linearly dependent on the layer depth \cite{sakai2024linearly,montanez2024towards}. Specifically, we optimized the four parameters, \(\alpha_I, b_I, \alpha_z, b_z\), appearing in Eqs.~\eqref{eq:gamma_linear} and \eqref{eq:beta_linear}. For the optimization, we employed Bayesian optimization using Optuna \cite{akiba2019optuna}.

The results are shown in Table~\ref{tab:resQAOA_linear}. When the number of atoms in the cost function optimization matches the number of atoms in QAOA sampling, the success rate improves across all layer depths for systems with 12, 16, and 24 atoms compared to the fixed-angle initial values. This result is expected and confirms that optimizing only the four parameters enables QAOA to work effectively, as demonstrated in previous studies \cite{sakai2024linearly,montanez2024towards}. However, for the transfer of rotation angle parameters optimized for smaller systems to QAOA sampling of larger systems, the results are similar to those in the previous subsection. Specifically, parameter transfer works well when parameters optimized for 12 atoms are applied to 16 atoms and when parameters optimized for 16 atoms are applied to 24 atoms. However, for 32-atom QAOA sampling, success rates remain low in all cases, and the fixed-angle initial values yield higher success rates.

These fixed-angle initial values, as mentioned right after Eq.~\eqref{eq:beta_linear}, were chosen to work well in fully connected Ising models with random coefficients for approximately 10 qubits. They were not tailored specifically for the energy functions derived from the cluster expansion method in this study. However, our results suggest that parameters effective for generic problems may sometimes yield better rotation angles than transferring parameters optimized for specific problems.

To further explore this, we focused on searching for better rotation angle parameters in the vicinity of the fixed-angle initial values by fixing \(\beta\) and optimizing only \(\gamma\) (optimizing only the two parameters \(\alpha_I, b_I\)). The results are presented in Table~\ref{tab:resQAOA_linear2}. For \(p = 5\), transferring rotation angles optimized for the 24-atom cost function to the 32-atom sampling achieved a success rate of nearly 10\%. While not very high, this represents an improvement over previous methods for smaller \(p\). Additionally, for \(p = 7\) and 8, transferring rotation angles optimized for 12-atom cost functions to 16-, 24-, and 32-atom systems resulted in moderate success rates, indicating that overfitting in the 12-atom optimization was effectively mitigated.

Finally, deviating from the idea of transferring parameters optimized for smaller systems to larger systems, we tested direct optimization of \(\gamma\) for the 32-atom cost function. This was relatively feasible and the resulting rotation angles were used for QAOA sampling. These results are reported in Table~\ref{tab:resQAOA_linear3} for reference. In all cases, the success rate for 32-atom QAOA sampling exceeded 10\%. Furthermore, applying these optimized parameters to 12-, 16-, and 24-atom systems yielded success rates of nearly 10\% or higher. Notably, for \(p = 5, 6, 7\), the success rates surpassed those obtained with fixed-angle initial values. 

However, for \(p = 8, 9\), fixed-angle initial values still outperformed in some cases, suggesting that there is significant room for improvement and further research in the selection of fixed-angle parameters.

\section{Conclusion}  \label{sec:conclusion}

In this study, we analyzed the elemental configuration optimization problem for the Au-Cu binary alloy using QAOA and demonstrated its effectiveness. Specifically, we showed that the optimal solution can be obtained for crystal systems containing up to 32 atoms, using the same number of qubits, as confirmed by quantum circuit simulations.  This work represents the first application of QAOA to a concrete materials structure optimization problem and holds significant importance as an extension of QAOA Hamiltonians to real-valued QUBO coefficients, rather than being limited to integer values. The study constitutes a novel challenge not extensively addressed in the existing literature.

We successfully solved problems involving 12, 16, and 24 atoms with QAOA by optimizing the rotation angles individually, using 5 to 9 layers in the ansatz. Moreover, for the 32-atom system, we identified effective fixed-angle parameters. On the other hand, it became evident that transferring rotation angles optimized for smaller systems to larger systems often led to overfitting. These results provide important insights for improving both the efficiency and accuracy of QAOA-based sampling.

For future prospects, one avenue of investigation is to extend the Ising Hamiltonian, which in this study was limited to a second-order cluster expansion, to include third- and fourth-order terms commonly used in standard cluster expansion methods. Additionally, in materials science, there is often a need to find optimal solutions under constraints such as fixed composition ratios (e.g., the Au:Cu ratio in this study). Exploring whether QAOA can handle such constrained optimization problems is another important direction.

The findings of this study lay the foundation for analyzing larger systems using real quantum computers. 
With the expected advancement of quantum hardware, it is anticipated that large-scale QAOA implementations will become feasible and may offer advantages over classical optimization methods. Future work is expected to contribute to the practical development of materials using quantum computers by exploring the applicability of QAOA to a wider variety of material systems and by improving rotation angle optimization methods.

\vspace{-0.1cm}
\begin{acknowledgments}
KF is supported by MEXT Quantum Leap Flagship Program (MEXT Q-LEAP) Grant No.~JPMXS0120319794, and JST COI-NEXT program Grant
No.~JPMJPF2014. 
K.N.O. is also supported by MEXT Quantum Leap Flagship Program (MEXT Q-LEAP) Grant No.~JPMXS0120319794.
\end{acknowledgments}

\bibliography{references}
\end{document}